\documentclass[showpacs,twocolumn,superscriptaddress]{revtex4}
\usepackage{doi}%<----------
\usepackage{hyperref}
\hypersetup{
  colorlinks=true,        % false: boxed links; true: colored links
  linkcolor=blue,         % color of internal links
  citecolor=cyan,         % color of links to bibliography
}

\usepackage{graphicx}% Include figure files
\usepackage{dcolumn}% Align table columns on decimal point
\usepackage{bm}% bold math
\usepackage{color}
\usepackage{enumitem}
\usepackage{amsmath}
\usepackage{amssymb}
\usepackage{orcidlink}
\usepackage{subcaption}
\usepackage{enumitem}
\usepackage{graphicx} % Required for inserting images

\usepackage{xcolor}

\begin{document}
\title{Photon Rings and Mass-Independent Observables of a Five-Dimensional\\ Charged AdS Black Hole
}

\author{Amrillokhon Mamatkhonov}
\email{mamatxonovamrillo@gmail.com}
\affiliation{National University of Uzbekistan, Tashkent 100174, Uzbekistan }

\author{Javokhir Sharipov}
\email{javohirsh100@gmail.com}
\affiliation{Institute of Fundamental and Applied Research, National Research University TIIAME, Kori Niyoziy 39, Tashkent 100000, Uzbekistan}
\affiliation{University of Tashkent for Applied Sciences, Str. Gavhar 1, Tashkent 100149, Uzbekistan}

\author{Tursunali Xamidov}
\email{xamidovtursunali@gmail.com}
\affiliation{Institute for Theoretical Physics and Cosmology, Zhejiang University of Technology, Hangzhou 310023, China}
\affiliation{Institute of Fundamental and Applied Research, National Research University TIIAME, Kori Niyoziy 39, Tashkent 100000, Uzbekistan}
\affiliation{Tashkent State Technical University, 100095 Tashkent, Uzbekistan}

\author{Chen Zhou}
\email{chenzhou@hit.edu.cn}
\affiliation{School of Physics, Harbin Institute of Technology, Harbin 150001, China}

\author{Sanjar Shaymatov}
\email{sanjar@astrin.uz}
\affiliation{Institute of Fundamental and Applied Research, National Research University TIIAME, Kori Niyoziy 39, Tashkent 100000, Uzbekistan}
\affiliation{National University of Uzbekistan, Tashkent 100174, Uzbekistan }

\author{Chengxun Yuan}
\email{yuancx@hit.edu.cn}
\affiliation{School of Physics, Harbin Institute of Technology, Harbin 150001, China}

\date{\today}

\begin{abstract}
We investigate the optical properties of a five-dimensional charged anti-de Sitter (AdS) black hole. Using the null geodesic equations, we determine the event horizon and photon sphere radii and the critical impact parameter, showing that all decrease with increasing charge, while the photon sphere radius remains independent of the AdS radius. 
To avoid dependence on the mass scale and the observer-dependent apparent size of the shadow in an AdS spacetime, we introduce two mass-independent geometric observables: the subring exponent $\gamma$ and the ratio of the critical impact parameter to that of the direct image of the innermost stable circular orbit (ISCO). We derive $\gamma$ in closed form and show that it is independent of the AdS radius and satisfies $2\pi/\sqrt{3}<\gamma\leq\pi\sqrt{2}$, above the values for four-dimensional Schwarzschild and Reissner--Nordstr\"om black holes.  
Using a Markov chain Monte Carlo (MCMC) forecast based on noiseless mock data with illustrative fractional uncertainties of $5\%$, we find that the dimensionless AdS radius can be constrained with a relative uncertainty of $16$--$19\%$, whereas the charge can be distinguished from zero only for the near-extremal fiducial model considered. Under these assumptions, the subring exponent provides a Gaussian separation of $3.4$--$5.9\sigma$ from the four-dimensional Schwarzschild black hole.
\end{abstract}

\maketitle

 \section{Introduction}

The study of black holes (BHs) has become an important part of modern gravitational physics. General Relativity (GR) provides a theoretical framework in which BHs can be described as solutions of Einstein’s field equations. However, observational studies of these objects have progressed rapidly only in recent decades. For example, gravitational-wave detections by LIGO \cite{Abbott16a, Abbott16b} and observations of black hole (BH) shadows with the Event Horizon Telescope (EHT) \cite{Akiyama19L1, Akiyama19L6, Akiyama22L12} have provided important evidence for their existence. These observations have also opened new possibilities for investigating their physical properties. Even with these developments, a number of questions concerning the nature and behavior of BHs remain unanswered.

BHs in higher-dimensional spacetimes have received increasing attention over the past few years, particularly in five-dimensional (5D) spacetimes. Such studies have been motivated by progress in braneworld models, string theory, and gauge/gravity duality, resulting in a number of remarkable theoretical findings~\cite{Horowitz2011}. In dimensions higher than four, the uniqueness theorems generally do not hold because the additional dimensions introduce extra degrees of freedom. The Schwarzschild–Tangherlini solution represents the five-dimensional vacuum BH that is both static and spherically symmetric in GR~\cite{Tangherlini1963}. Beyond this solution, supergravity and string theory contain various higher-dimensional charged BH solutions, including slowly rotating and charged forms~\cite{Aliev06,Chong05a,Cvetic96} extended to various astrophysical conditions \cite{Aliev07a,An18,Shaymatov20b,2022IJMPD..3150120D,Shaymatov24PRD.110d4042S}. The higher-dimensional Reissner–Nordström (RN) BH has been extensively examined in Refs.~\cite{Aman06,Xie22PLB,Sharif16,Xamidov2025EPJP}, with particular attention to its accretion properties, thermodynamic curvature, and scalarization.

The properties of BHs can be further explored through photon geodesics, as their trajectories provide valuable information about the underlying spacetime geometry and dynamics. In particular, the ray-tracing method can be used to investigate the observable features arising from photon trajectories~\cite{Gralla_2019,Peng_2021,SHARIPOV2026128}. It provides a standard framework for probing the geodesic structure of spacetime and examining how black hole parameters influence photon trajectories, and has therefore been widely used in recent studies~\cite{Cui24,Chen2025,Ziqiang25,Liu:2021yev,Sharipov:2025yfw,Hu2023}. The strong gravitational fields associated with compact astrophysical objects provide valuable opportunities to examine the predictions of gravitational theories. In particular, the study of accretion disks through their structure~\cite{Chandrasekhar98,Faraji:2025,Bambi17e,Boos:2025nzc}, X-ray emission \cite{Bambi16b,Bambi12a}, and dynamical behavior provides valuable information about gravity in the strong-field regime~\cite{Abramowicz13,Auchettl17ApJ,Fender04mnrs}. Sgr A$^{\star}$ at the Galactic center is another important example for studying accretion disk and strong-field gravity \cite{you2024,Nucita07,Afrin23ApJ,Ghez05,Ghez00,Vagnozzi23EHT,Imdiker23CQG,Shaymatov2023,Collodel_2021,Alloqulov24CPC,Boshkayev21PRD,Alloqulov24EPJP,Nozari2025b,Nozari2025,Liu:2020vkh,Igata2025,Xamidov:2026hrt,XamidovAccretion2025,Sharipov:2026syz}. In the context of 5D spacetime, charged AdS BHs are of particular interest because their charge can alter geodesic motion and, consequently, the structure of surrounding environment around the BH.

Many studies compare theoretical BH shadow sizes with EHT observations~\cite{Akiyama19L1,Akiyama22L12}. However, such comparisons require a mass scale, usually inferred from observations within four-dimensional gravity. In five dimensions, the mass parameter has the dimension of length squared, making its relation to observationally inferred masses model-dependent. Moreover, in an AdS spacetime, the apparent angular size of the shadow depends on the observer's position. These considerations motivate the use of mass-independent geometric observables.

One such observable is the subring exponent, which characterizes how successive photon subrings approach the critical curve and is determined by the instability of the circular photon orbit~\cite{Johnson2020,Gralla_2019,Cardoso2009}. Future high-resolution VLBI facilities, such as the next-generation Event Horizon Telescope (ngEHT) and the proposed Black Hole Explorer (BHEX), may provide new opportunities to study these structures~\cite{Johnson2023ngEHT,Lupsasca2024BHEX}. Another observable is the ratio of the shadow radius to the radius of the direct image of the innermost stable circular orbit (ISCO), a ratio that is independent of the overall mass scale. In five dimensions, stable circular orbits are absent in the asymptotically flat Schwarzschild--Tangherlini spacetime~\cite{Hackmann2008}, whereas the AdS term allows such orbits to exist.

In this work, we investigate the photon dynamics and optical properties of a five-dimensional charged AdS black hole, described by the metric given in Ref.~\cite{Ali_2023}. We focus on photon trajectories, the two mass-independent geometric observables discussed above, and their potential to constrain the black hole parameters through a Bayesian MCMC forecast. The paper is organized as follows. In Sec.~\ref{Sec:II}, we study null geodesics, the characteristic radii, and the classification of photon trajectories. In Sec.~\ref{Sec:Obs}, we introduce the mass-independent observables and the forecast setup. The forecast results are presented in Sec.~\ref{Sec:Forecast}, and our conclusions are given in Sec.~\ref{Sec:Conc}.

\section{Spacetime metric and null geodesics}\label{Sec:II}

This section is devoted to the analysis of null geodesics for the 5D case of the higher-dimensional charged AdS BH. The full metric is expressed in the form \cite{Ali_2023}:
\begin{equation}
    ds^2 = -f(r)dt^2 + f(r)^{-1}dr^2 + r^2d\Omega_{D-2}^2 , \label{ds2}
\end{equation}
with
\begin{equation}
    f(r) = 1 - \frac{2M_D}{r^{D-3}} + \frac{Q_D^2}{r^{2D - 6}}+\frac{r^2}{l^2} , \label{fr}
\end{equation}
\begin{eqnarray}
d\Omega_{D-2}^{2}
=&&
d\theta_1^{2}
+
\sin^{2}\theta_1\,d\theta_2^{2}
+\cdots\nonumber\\&&+
\left(
\prod_{j=1}^{D-3}\sin^{2}\theta_j
\right)
d\theta_{D-2}^{2},\label{metricOmega}
\end{eqnarray}
where $M_D$ denotes the mass, whereas $Q_D$ represents the charge in higher dimensions, and $l$ corresponds to the AdS radius. $d\Omega_{D-2}^2$ specifies the line element associated with the (D-2)-dimensional unit sphere.
The parameters $M_D$ and $Q_D$ are related to the ADM mass $\mathcal{M}$ and the electric charge $\mathcal{Q}$ through $M_D=8\pi\mathcal{M}/[(D-2)\Omega_{D-2}]$ and $Q_D=8\pi\mathcal{Q}/[(D-2)\Omega_{D-2}]$, where $\Omega_{D-2}=2\pi^{(D-1)/2}/\Gamma\!\left(\frac{D-1}{2}\right)$ is the area of the unit $(D-2)$-sphere~\cite{Ali_2023}. For $D=5$, $\Omega_3=2\pi^2$ and hence $\mathcal{M}=3\pi M_5/4$ (in units $G_5=1$). Note that $M_5$ and $Q_5$ have dimensions of length squared. In what follows, we drop the subscript, $M\equiv M_5$ and $Q\equiv Q_5$, and set $M=1$, i.e., all lengths are measured in units of $\sqrt{M_5}$.

For five dimensions, Eq.~\eqref{metricOmega}  takes the form below:
\begin{equation}
\label{metric}
d\Omega_{3}^2=d\theta^2+\sin^2\theta\left(d\chi^2+\sin^2\chi\, d\phi^2\right).
\end{equation}
For the motion of massless particles in a charged AdS BH spacetime, the Lagrangian can be expressed in the form~\cite{1973grav.book.....M,1983mtbh.book.....C}:
\begin{equation}
\mathcal{L}=\frac{1}{2}\,g_{\mu\nu}\,\frac{dx^{\mu}}{ds}\,\frac{dx^{\nu}}{ds}.
\end{equation}
Here, $s$ denotes the affine parameter, while the photon momentum takes the form
\begin{equation}
    p^{\alpha}=\frac{dx^{\alpha}}{ds}\, .
\end{equation}
Since the metric possesses Killing vectors associated with the $t$ and $\phi$ coordinates, meaning that its components do not explicitly depend on these coordinates, the corresponding constants of motion, namely the photon's energy $E=-p_t$ and angular momentum $L=p_{\phi}$, are conserved quantities along the geodesic. 
Owing to the spherical symmetry, we can choose a two-dimensional orbital plane by setting $\theta=\chi=\pi/2$. Using the null geodesic normalization condition ($g_{\mu\nu}\dot{x}^{\mu}\dot{x}^{\nu}=0$), the equations describing the photon's trajectory can be expressed as: 
\begin{figure*}[!htb]
    \centering
    \includegraphics[width=0.48\linewidth]{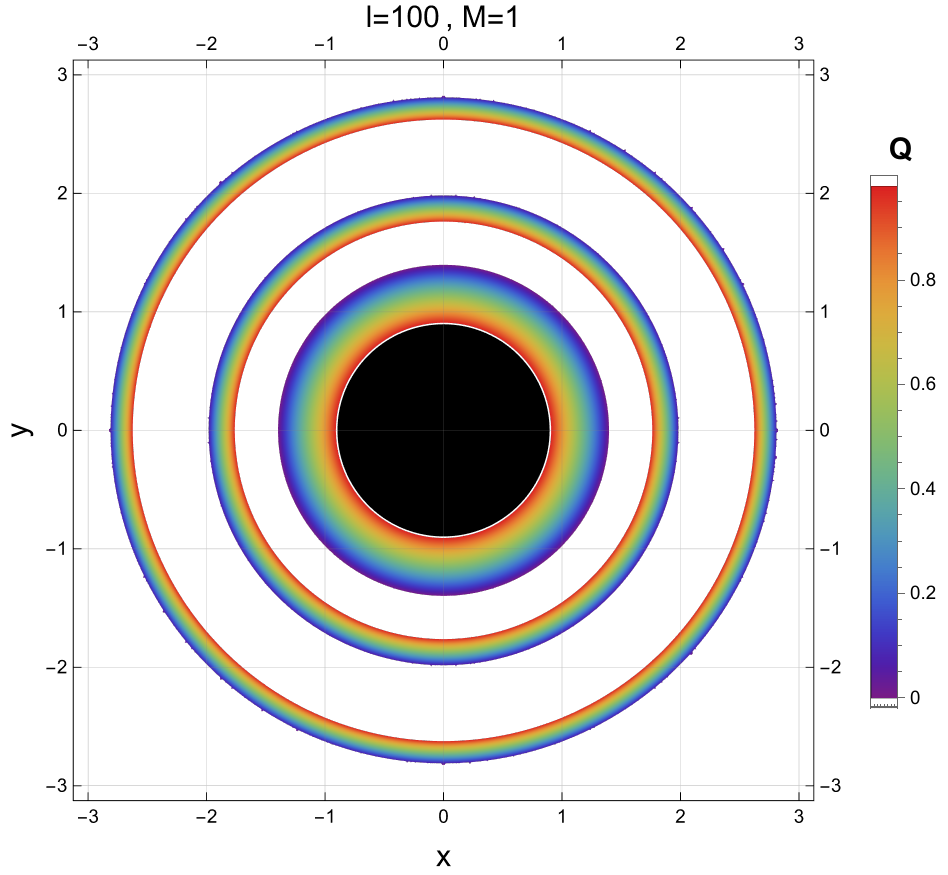}\hspace{0.5cm}
    \includegraphics[width=0.48\linewidth]{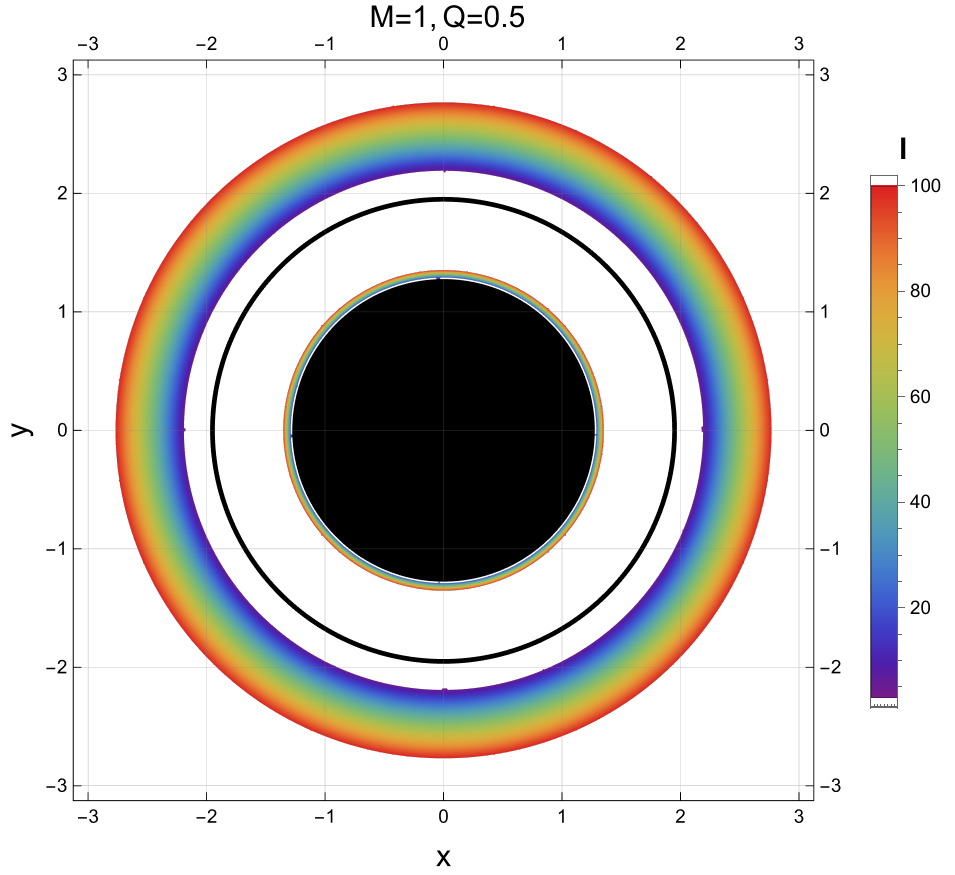}\hspace{0.5cm}
    \caption{The plots demonstrate the characteristic radii in the charged AdS BH spacetime for different values of the charge \(Q\) (left) and the AdS radius \(l\) (right). The circles provide a schematic comparison of the characteristic scales: $r_{\rm h}$ and $r_{\rm ph}$ are radial coordinates, whereas $R_{\rm sh}=b_c$ is an impact parameter.}
    
\label{fig:function_r}
\end{figure*}

\begin{figure*}
    \centering

    \begin{subfigure}{0.33\textwidth}
    % \label{fig: ray1}
        \centering
        \includegraphics[width=\linewidth,height=1\textwidth,keepaspectratio]{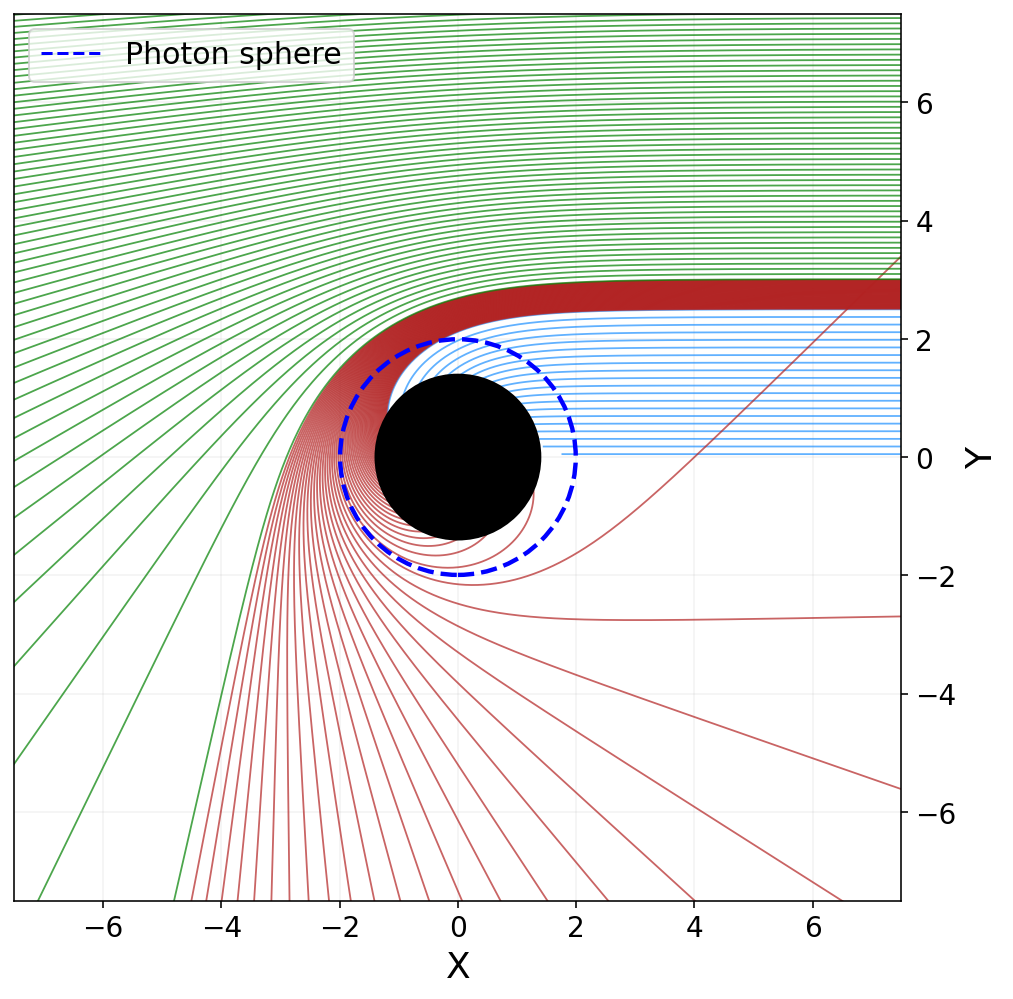}
        \caption{$Q=0.2$}
    \end{subfigure}
    \hfill
    \begin{subfigure}{0.3275\textwidth}
        \centering
        \includegraphics[width=\linewidth,height=1\textwidth,keepaspectratio]{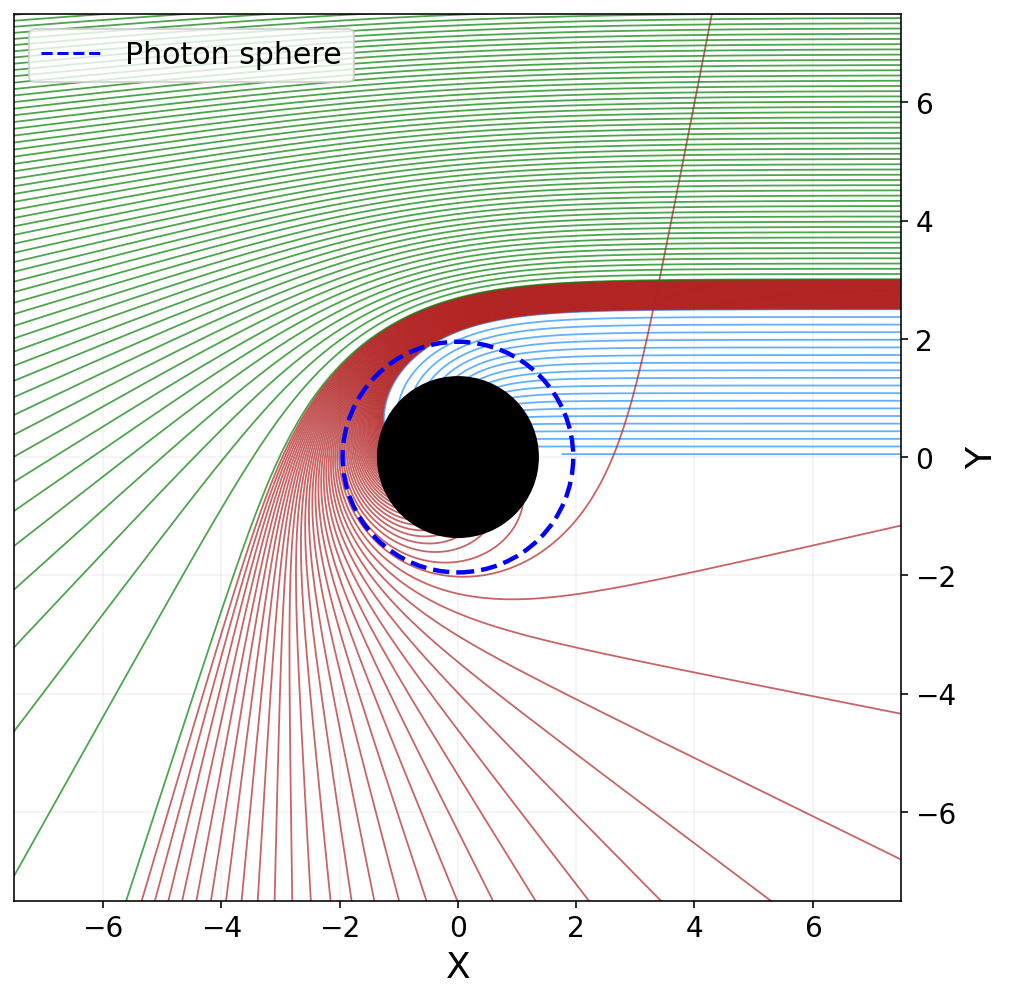}
        \caption{$Q=0.5$}
    \end{subfigure}
    \hfill
    \begin{subfigure}{0.33\textwidth}
        \centering
        \includegraphics[width=\linewidth,height=1\textwidth,keepaspectratio]{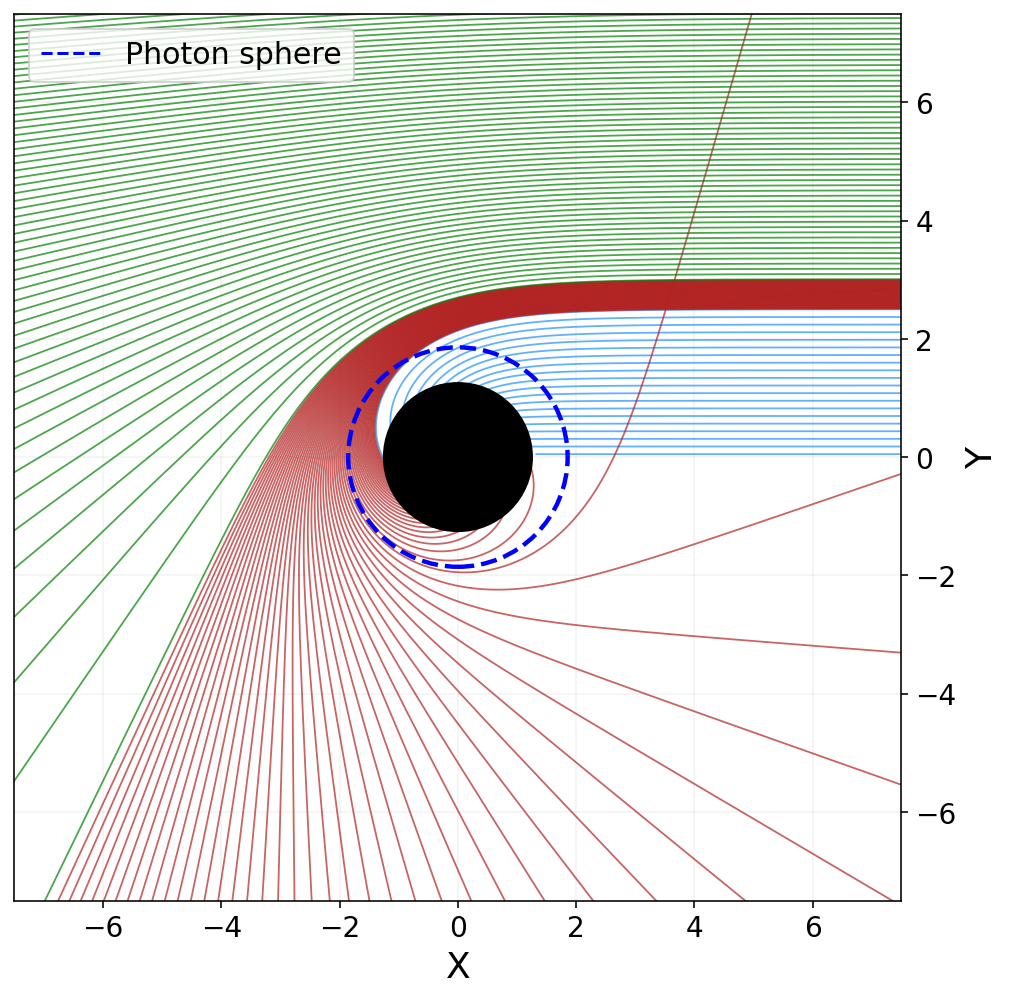}
        \caption{$Q=0.8$}
    \end{subfigure}

    \caption{The photon trajectories are presented in separate plots for various values of $Q$. The trajectories are colored blue, red, and green for $0<b<2.5$, $2.5<b<3$, and $b>3$, respectively, with $l=10$ and $M=1$. 
}
    
\label{b_func}
\end{figure*}

\begin{figure*}
    \centering

    \begin{subfigure}{0.33\textwidth}
        \centering
\includegraphics[width=\linewidth,height=1\textwidth,keepaspectratio]{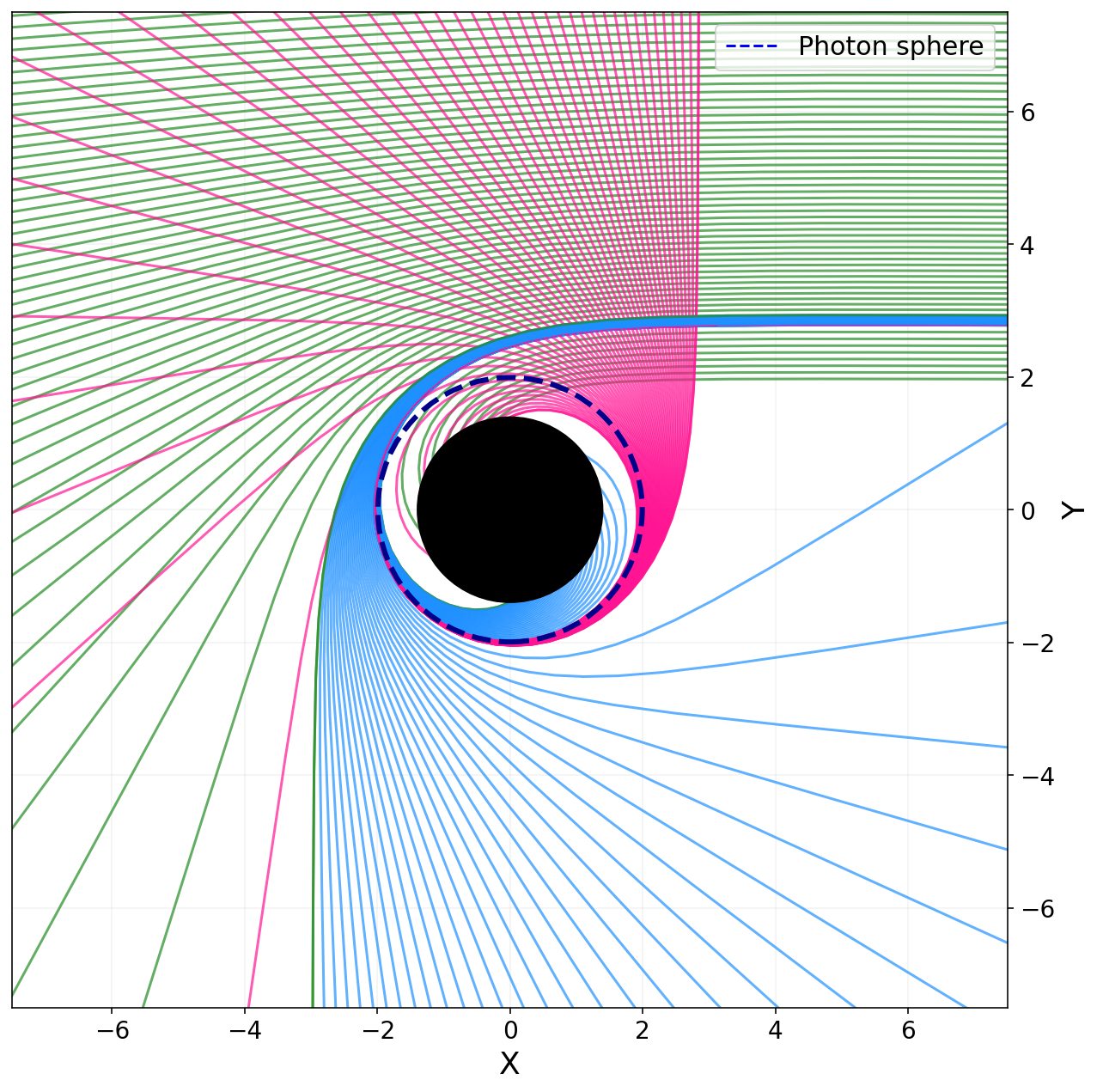}
        \caption{$Q=0.2$}
    \end{subfigure}
    \hfill
    \begin{subfigure}{0.325\textwidth}
        \centering
        \includegraphics[width=\linewidth,height=1\textwidth,keepaspectratio]{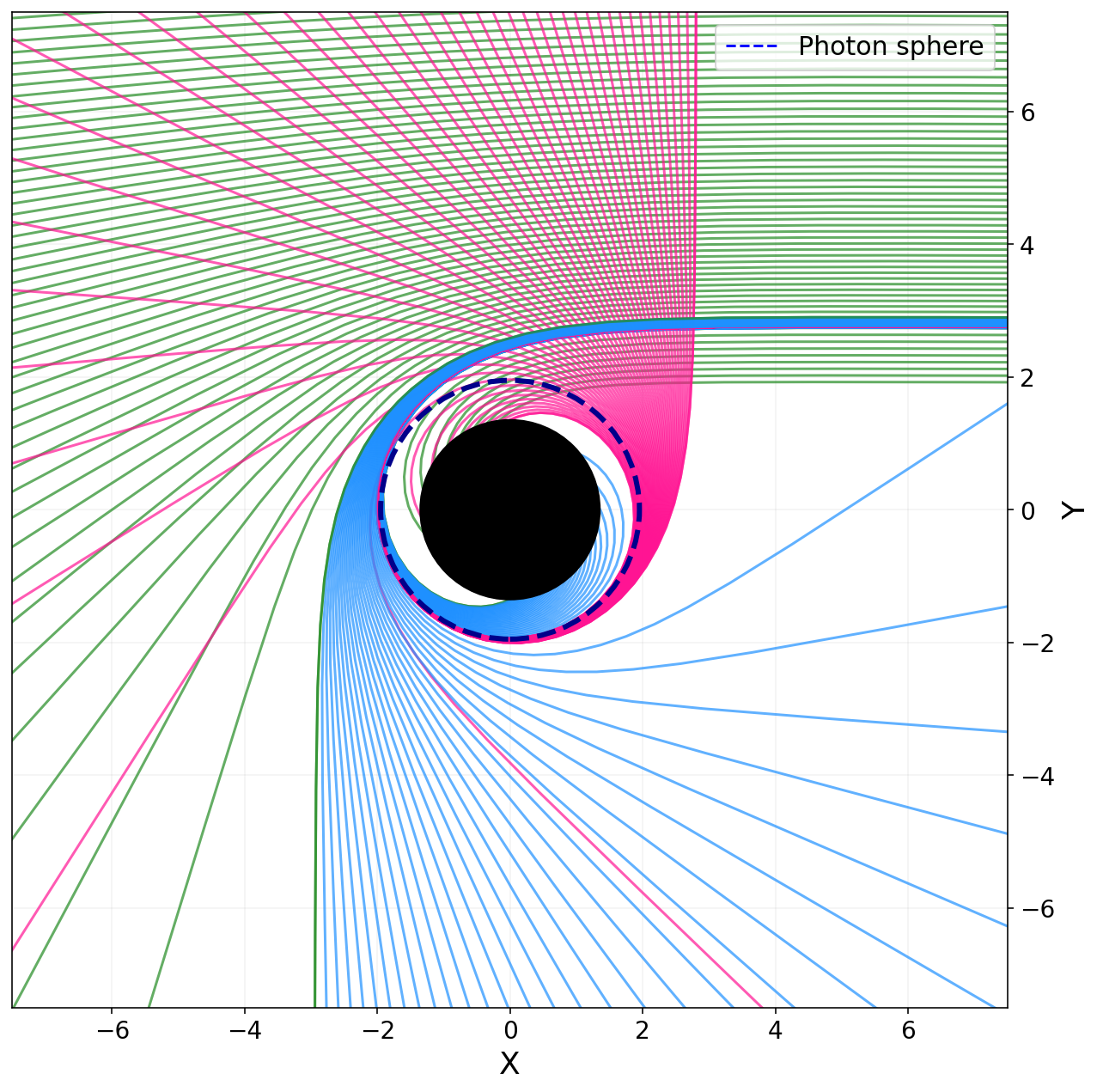}
        \caption{$Q=0.5$}
    \end{subfigure}
    \hfill
    \begin{subfigure}{0.33\textwidth}
        \centering
        \includegraphics[width=\linewidth,height=1\textwidth,keepaspectratio]{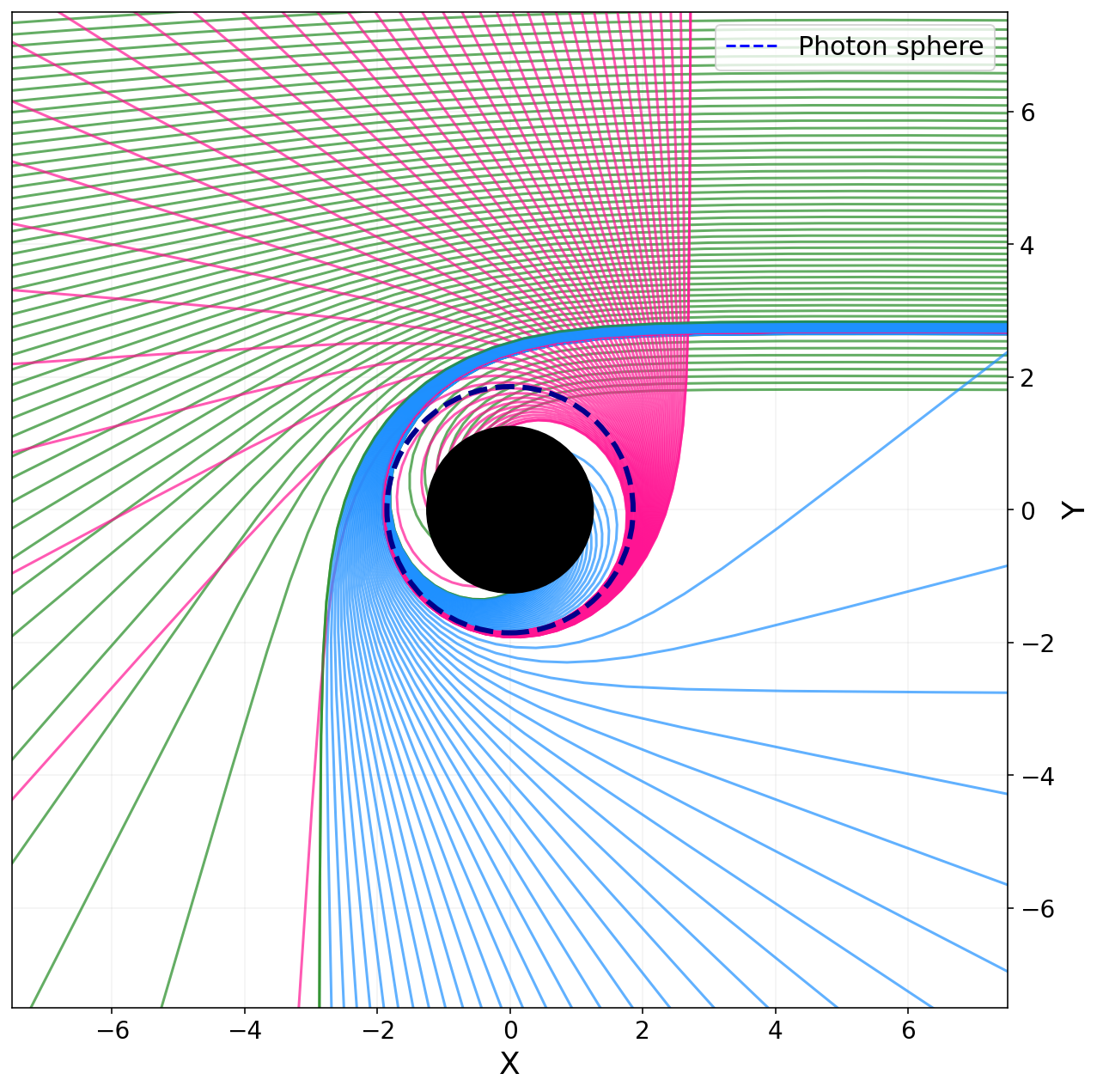}
        \caption{$Q=0.8$}
    \end{subfigure}

    \caption{ Ray-tracing results in the vicinity of a charged AdS BH are displayed. The trajectories are classified according to the number of revolutions $\eta$ as $1/4<\eta<3/4$, $3/4<\eta<5/4$, and $\eta>5/4$, corresponding to the green, blue, and magenta curves, respectively, with $l=10$ and $M=1$.
    }
   
\label{ray1}
\end{figure*}
\begin{figure}
    \centering
    \includegraphics[scale=0.38]{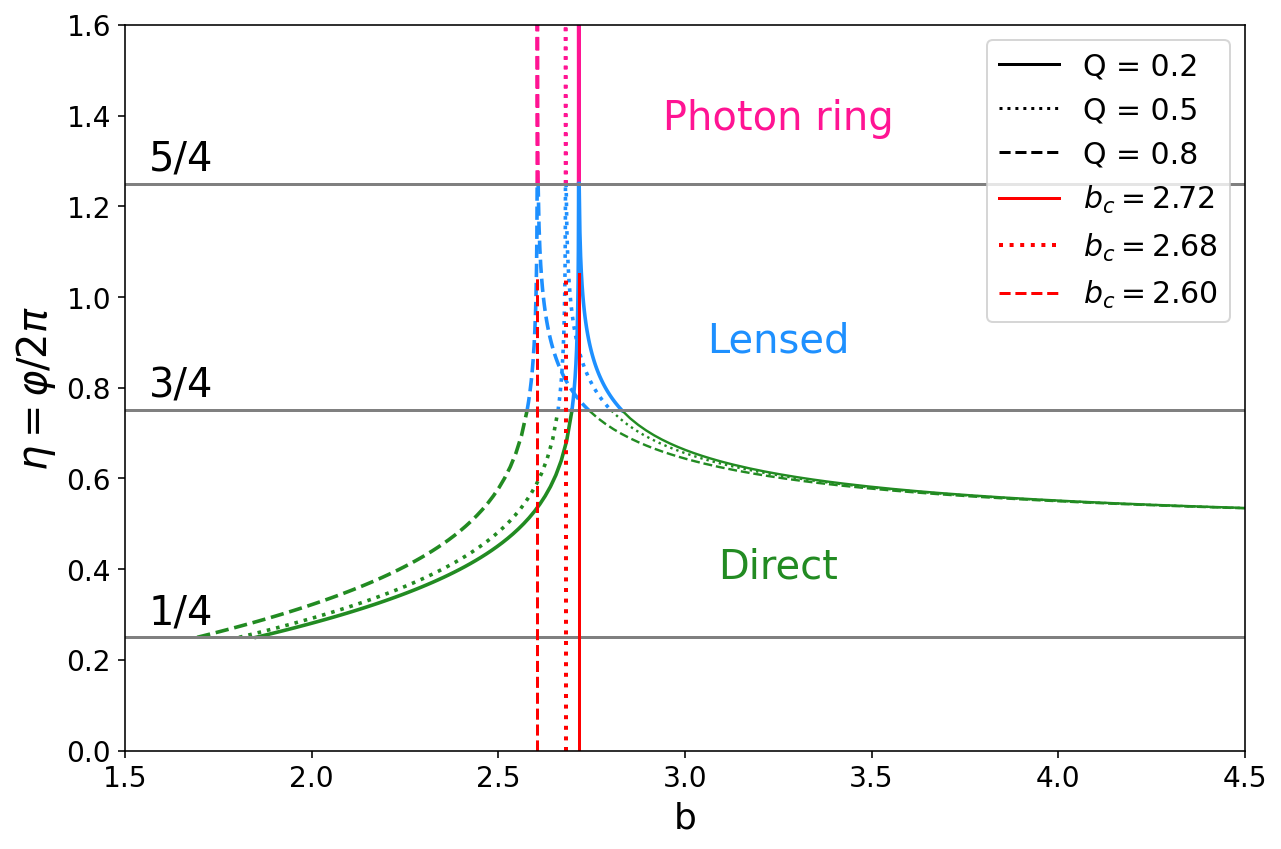}
    \caption{ The number of revolutions $\eta$ is plotted against the impact parameter $b$ for three different values of the charge parameter Q, with $l=10$.
    }
\label{F_b1}
\end{figure}

\begin{equation}
    \frac{dt}{ds}=\frac{E}{f(r)}, \quad
    \frac{d\phi}{ds}=\frac{L}{r^2}\ ,
\end{equation}

\begin{equation}
    \left(\frac{dr}{ds}\right)^2=E^2-f(r)\frac{L^2}{r^2}=E^2-V_{\rm eff}(r)\ .
\end{equation}

Here, $V_{\mathrm{eff}}$ denotes the effective potential governing the radial motion of a massless particle. The radius of the photon sphere $r_{\rm ph}$ is found by applying the extremum condition
$\left.\frac{dV_{\mathrm{eff}}}{dr}\right|_{r=r_{\mathrm{ph}}}=0$, which yields $r_{\rm ph}^2=2M+\sqrt{4M^2-3Q^2}$. The photon sphere radius is thus independent of the AdS radius $l$, and it exists for $Q\leq 2M/\sqrt{3}$.

After rescaling the affine parameter according to $s'=Ls$ and defining the impact parameter as $b=L/E$, the resulting equations take the following form:
\begin{equation}
    \frac{dt}{ds'}=\frac{1}{bf(r)}, \quad
    \frac{d\phi}{ds'}=\frac{1}{r^2}\ \label{radial1},
\end{equation}
\begin{equation}
    \left(\frac{dr}{ds'}\right)^2=\frac{1}{b^2}-\frac{f(r)}{r^2}\equiv F(r)\ .
    \label{radialM}
\end{equation} 

Substituting $r=r_{\rm ph}$ into $F(r)=0$ allows us to obtain the critical impact parameter $b_c$~\cite{Perlick2022,Sharipov:2025yfw}

\begin{equation}
R_{sh}=b_c = \frac{r_{\rm ph}}{\sqrt{f(r_{\rm ph})}}\ .
\end{equation}
In an asymptotically flat spacetime, $b_c$ coincides with the shadow radius measured by a distant observer. The AdS spacetime, however, is not asymptotically flat: since $f(r)/r^2\to 1/l^2$ as $r\to\infty$, Eq.~\eqref{radialM} gives $F(r\to\infty)=1/b^2-1/l^2$, so that only photons with $b<l$ reach the asymptotic region. For a static observer located at $r_O$, the angular radius of the shadow is 
\begin{equation}
\sin\alpha_{\rm sh}=\frac{b_c\sqrt{f(r_O)}}{r_O}\ ,
\label{alphash}
\end{equation}
which tends to $b_c/l$ as $r_O\to\infty$. Throughout this work, we construct the images in the plane of the impact parameter and, in this sense, refer to $R_{\rm sh}\equiv b_c$ as the shadow radius.

Figure~\ref{fig:function_r} shows the behavior of the characteristic radii for different values of the charge and AdS radius. In the left panel, the concentric colored circles represent the shadow, photon sphere, and event horizon, ordered from the largest to the smallest radius. 
The black disk indicates the event horizon for the largest charge considered, which must remain below the extremal value $Q_{\rm ext}\simeq0.99995$ for $l=100$. As the charge $Q$ increases, each of these radii shifts toward smaller values. The right panel is obtained for $Q=0.5$, while $l$ is varied from $3$ to $100$. The photon sphere radius remains unchanged because it is independent of $l$ and is therefore represented by a black circle. In contrast, larger values of $l$ result in larger event horizon and shadow radii.

Applying Eq.~\eqref{radial1} and the substitution $r=1/u$, we rewrite Eq.~\eqref{radialM} as follows~\cite{Sharipov:2026syz,XamidovAccretion2025}:

\begin{equation}
    \left( \frac{du}{d\phi} \right)^2 = \frac{1}{b^2} - u^2 f\left(\frac{1}{u} \right)\equiv G(u)\ .
\label{Gu}
\end{equation}

The total azimuthal angle $\varphi$ swept by the photon (for scattered rays the deflection angle is $\varphi-\pi$) can be obtained by transforming the above differential equation into an integral form. In the resulting expression, the first integral corresponds to photons captured by the BH ($b<b_c$), whereas the second integral corresponds to photons that escape to infinity after being deflected by the gravitational field; in the AdS case, ``infinity'' means the asymptotic region, which is reached only for $b<l$, since $G(u\to0)=1/b^2-1/l^2$ \cite{SHARIPOV2026128,Sharipov:2025yfw}.

\begin{equation}\label{phiint}
\varphi = 
\begin{cases} 
\displaystyle \ \ \
\int\limits_0^{u_{\rm h}} \frac{du}{\sqrt{G(u)}}\ ,  \quad & b < b_c\  \\[12pt]
\displaystyle
2 \int\limits_0^{u_{\min}} \frac{du}{\sqrt{G(u)}}\ , \quad & l>b > b_c\ .
\end{cases}
\end{equation}
Here, $u_{\rm h}=1/r_{\rm h}$ represents the inverse of the event horizon radius. For $l>b>b_c$, $u_{\min}$ corresponds to the closest approach of the photon to the BH and is determined by the smallest positive solution of $G(u)=0$.

Photons traveling in parallel from infinity undergo different deflection angles in the vicinity of the BH, depending on their impact parameter. This behavior is shown in Fig.~\ref{b_func}, which illustrates the photon trajectories near the BH. The colors in the figure correspond to different intervals of $b$: 
blue for $0<b<2.5$, red for $2.5<b<3$, and green for $b>3$. For the parameter sets $(Q, M,l) = (0.2, 1, 10)$, $(0.5, 1, 10)$, and $(0.8, 1, 10)$, the resulting critical impact parameters are found to be $b_c = 2.7153$, $2.6802$, and $2.6046$, respectively. The central black region in the figure denotes the event horizon, while the blue dashed ring marks the photon sphere; both radii shrink as the BH charge parameter increases. As the charge parameter is reduced, the red rays undergo stronger bending around the BH.

Photon trajectories in the vicinity of a BH can be classified into three distinct types (see, e.g., \cite{Gralla_2019,Peng_2021,SHARIPOV2026128}). Their boundaries are determined using the number of revolutions ($\eta=\frac{\varphi}{2\pi}$)~\cite{Sharipov:2025yfw,XamidovAccretion2025}:
\begin{eqnarray}
\eta=\frac{2N-1}{4},\qquad N=1,2,3,\cdots.
\end{eqnarray}
\begin{itemize}
\item 1st type, direct trajectories: $\frac{1}{4}<\eta<\frac{3}{4}$, with the corresponding impact-parameter ranges $(b_1^ {-},b_2^ {-})\cup$
$(b_2^{+},l)$, indicated by green curves in 
Fig.~\ref{ray1}.
\item 2nd type, lensed trajectories: $\frac{3}{4}<\eta<\frac{5}{4}$, $(b_2^ {-},b_3^ {-})\cup(b_3^ {+},b_2^{+})$, blue curves.
\item 3rd type, photon-ring trajectories: $\eta>\frac{5}{4}$, $(b_3^ {-},b_3^ {+})$, magenta curves.
\end{itemize}

\begin{table}
\centering
\renewcommand{\arraystretch}{1.3}
\resizebox{0.48\textwidth}{!}{
\begin{tabular}{|c|c|c|c|c|c|c|c|c|}
\hline
\textbf{BHs} & \textbf{$r_{\rm h}$} & \textbf{$r_{\rm ph}$} & \textbf{$b^-_1$} & \textbf{$b^-_2$} & \textbf{$b^-_3$} & \textbf{$R_{sh}$} & \textbf{$b^+_3$} & \textbf{$b^+_2$} \\
\hline
\textbf{$Q = 0.2$} & 1.393 & 1.992 & 1.851 & 2.697 & 2.7150 & 2.7153 & 2.717 & 2.832 \\
\hline
\textbf{$Q = 0.5$} & 1.353 & 1.950 & 1.805 & 2.660 & 2.6799 & 2.6802 & 2.682 & 2.803 \\
\hline
\textbf{$Q = 0.8$} & 1.252 & 1.855 & 1.693 & 2.577 & 2.6040 & 2.6046 & 2.607 & 2.744 \\
\hline
\end{tabular}
}
\caption{The values of the event horizon $r_{\rm h}$, photon sphere radius $r_{\rm ph}$, impact parameter $b_n^ \pm$, and shadow radius $R_{sh}$ for several charge parameters $Q$ of the charged AdS BH, with $l=10$. 
}
\label{table1}
\end{table}

The results in Table~\ref{table1} are consistent with those shown in Figs.~\ref{ray1} and~\ref{F_b1}. Figure~\ref{F_b1} shows the variation of the total number of revolutions $\eta$ with the impact parameter $b$ for three values of the charge parameter $Q$. The horizontal lines indicate $\eta=1/4$, $3/4$, and $5/4$, and their intersections with the curves define the corresponding values $b_n^\pm$. Near the critical impact parameter $b_c$, the number of revolutions $\eta$ becomes very large. As the charge parameter $Q$ increases, the critical impact parameter $b_c$ decreases, as seen in Fig.~\ref{F_b1}. The corresponding charged AdS BH quantities, including $r_{\rm h}$, $r_{\rm ph}$, $b_n^\pm$, and the shadow radius $R_{\rm sh}$, are listed in Table~\ref{table1} for different values of $Q$. The table shows that all characteristic quantities  decrease as $Q$ increases. In the limit $b\to b_c$, the number of revolutions satisfies $\eta\to\infty$.

\section{Mass-independent observables}\label{Sec:Obs}

The shadow radius $b_c$ obtained in Sec.~\ref{Sec:II} is measured in units of $\sqrt{M}$. In five dimensions, however, $M$ has the dimension of length squared, and there is no direct way to relate it to the mass of an observed black hole, which is usually inferred from the motion of nearby stars or gas using four-dimensional gravity. Moreover, in the AdS spacetime the apparent size of the shadow depends on the position of the observer (see Eq.~\eqref{alphash}). A direct comparison of $b_c$ with the shadow sizes measured by the EHT would therefore rely on arbitrary assumptions. In this section, we instead consider two dimensionless quantities that do not depend on the mass scale and are insensitive to the position of the observer. Both are determined by the geometry of the spacetime alone and do not require a model for the emission of the disk. Since the metric depends on the charge only through $Q^2$, the sign of $Q$ does not affect the photon trajectories. We therefore characterize the black hole by the dimensionless charge $e=|Q|/M$ and the dimensionless AdS radius $\lambda=l/\sqrt{M}$.

\subsection{Photon subrings}

Light rays that pass close to the photon sphere can circle the black hole several times before reaching the observer. As a result, the image contains a sequence of thin rings, the photon subrings, labeled by the number $n$ of half-orbits around the black hole. Since the circular photon orbit is unstable, neighboring rays separate exponentially. Near the critical impact parameter, the total change in the azimuthal angle along a ray grows logarithmically, $\Delta\phi\simeq-(\Omega_c/\lambda_L)\ln|b-b_c|+{\rm const}$. As a result, successive images, which differ by $\Delta\phi=\pi$, approach the critical curve geometrically
~\cite{Johnson2020,Gralla_2019},
\begin{equation}
\frac{|b_{n+1}-b_c|}{|b_n-b_c|}\simeq \exp(-\gamma)\ ,\qquad \gamma=\frac{\pi\lambda_L}{\Omega_c}\ .
\label{subring}
\end{equation}

Here, $\Omega_c=\sqrt{f(r_{\rm ph})}/r_{\rm ph}$ is the angular velocity of the photon on the circular orbit, and $\lambda_L$ is the Lyapunov exponent of this orbit. Thus, each subring is narrower than the previous one by the factor $\exp(-\gamma)$. For optically thin emission and $n\gg1$, the same factor approximately describes the flux ratio of successive subrings~\cite{Johnson2020}; here, however, we use $\gamma$ only as a geometric observable. For a static, spherically symmetric metric of the form~\eqref{ds2}, $\lambda_L^2=f\left(2f-r^2f''\right)/(2r^2)$ at $r=r_{\rm ph}$~\cite{Cardoso2009}, which gives
\begin{equation}
\gamma=\pi\sqrt{f-\frac{r^2f''}{2}}\,\Bigg|_{r=r_{\rm ph}} .
\label{gammadef}
\end{equation}

The term $r^2/l^2$ in $f(r)$ does not contribute to the combination $f-r^2f''/2$, and $r_{\rm ph}$ does not depend on $l$. Therefore, $\gamma$ is independent of the AdS radius. Using the photon-sphere condition $r_{\rm ph}^4-4Mr_{\rm ph}^2+3Q^2=0$, Eq.~\eqref{gammadef} reduces to
\begin{equation}
\gamma=2\pi\sqrt{1-\frac{2M}{r_{\rm ph}^2}}=2\pi\sqrt{\frac{\sqrt{4-3e^2}}{2+\sqrt{4-3e^2}}}\ .
\label{gammaclosed}
\end{equation}
For the Schwarzschild--Tangherlini black hole ($e=0$), $\gamma=\pi\sqrt{2}\simeq4.44$, and $\gamma$ decreases monotonically with the charge, approaching $2\pi/\sqrt{3}\simeq3.63$ as $e\to1$, i.e., in the asymptotically flat extremal limit. Thus, $\gamma$ is always larger than its value $\gamma=\pi$ for the 4D Schwarzschild black hole, as well as its values $\pi/\sqrt{2}\leq\gamma\leq\pi$ for the 4D Reissner--Nordstr\"om black hole, i.e., the subrings of the 5D black hole become narrower faster (see Table~\ref{tab:gamma}). For small charges, $\gamma\simeq\pi\sqrt{2}\,(1-3e^2/32)$, and $\gamma$ is therefore only weakly sensitive to $e$ near $e=0$. Finally, since $\gamma$ is the ratio of two rates measured with the same time coordinate, it does not depend on the position of the observer, which is important in the AdS spacetime. 
We also note that the local sky of a five-dimensional observer is three-dimensional, and the critical curve is a two-sphere rather than a circle. Owing to the spherical symmetry, however, the radial position of any image on the sky is determined by its impact parameter $b$ alone, which we use throughout.

\begin{table}[t]
\centering
\small
\setlength{\tabcolsep}{4pt}
\renewcommand{\arraystretch}{1.3}
\begin{tabular}{|c|c|c|c|c|c|}
\hline
 & $e=0$ & $e=0.5$ & $e=0.8$ & $e=0.99$ & 4D Schw. \\
\hline
$\gamma$ & 4.443 & 4.326 & 4.067 & 3.663 & 3.142 \\
\hline
$\exp(-\gamma)$ & 0.0118 & 0.0132 & 0.0171 & 0.0257 & 0.0432 \\
\hline
\end{tabular}
\caption{The subring exponent $\gamma$ and the geometric contraction factor $\exp(-\gamma)$ of successive photon subrings for the 5D charged AdS BH (independent of $l$), compared with the 4D Schwarzschild BH.}
\label{tab:gamma}
\end{table}

\subsection{Inner edge of the disk image}

We now consider an idealized, geometrically thin emitting region in the equatorial plane whose inner edge coincides with the ISCO. This assumption specifies only the geometry of the source and not an accretion-flow or radiation-transport model. The position of the inner edge in the image then depends only on the ISCO radius and on the photon trajectories.

Using the standard expressions for timelike circular orbits~\cite{Shapiro83,Shaymatov22a}, we find that, for the metric function~\eqref{fr} with $D=5$, the specific angular momentum of a circular orbit in the equatorial plane is
\begin{equation}
\tilde L^2=\frac{2M-2Q^2/r^2+r^4/l^2}{1-4M/r^2+3Q^2/r^4}\ ,
\label{Lcirc}
\end{equation}
and the ISCO corresponds to its minimum, $d\tilde L/dr=0$. With $y=r^2$, this condition reads
\begin{equation}
\begin{split}
y^3\left(y^2-6My+6Q^2\right)={}&l^2\big[(4M^2-Q^2)\,y^2\\
&-6MQ^2y+3Q^4\big].
\end{split}
\label{iscoeq}
\end{equation}
In the asymptotically flat limit $l\to\infty$, Eq.~\eqref{iscoeq} has no solution for $Q<M$, i.e., there are no stable circular orbits, as is known for the five-dimensional Schwarzschild--Tangherlini spacetime~\cite{Hackmann2008}. Hence, stable circular timelike geodesics, and thus the ISCO, exist here only because of the AdS term. For $Q=0$, Eq.~\eqref{iscoeq} gives $r_{\rm ISCO}^4(r_{\rm ISCO}^2-6M)=4M^2l^2$, which implies $r_{\rm ISCO}\propto l^{1/3}$ for large $l$.

For a face-on observer in the asymptotic region ($r_O\to\infty$), a photon emitted from the inner edge reaches the observer after turning by the angle $\Delta\phi=\pi/2$. From Eq.~\eqref{Gu}, the radius $b_{\rm ISCO}$ of the direct image of the inner edge is then determined by
\begin{equation}
\frac{\pi}{2}=\int_0^{1/r_{\rm ISCO}}\frac{du}{\sqrt{G(u;b_{\rm ISCO})}}\ ,
\label{bisco}
\end{equation}
where we write $G(u;b)$ to show its dependence on the impact parameter, and the photon moves outward without a turning point. We then define the observable
\begin{equation}
\mathcal{O}=\frac{R_{\rm sh}}{b_{\rm ISCO}}=\frac{b_c}{b_{\rm ISCO}}\ .
\label{Oratio}
\end{equation}
Both radii are measured in the same image plane, and the mass scale therefore cancels in their ratio. For a static observer at a finite radius $r_O$, the lower limit of the integral in Eq.~\eqref{bisco} becomes $1/r_O$, and $\mathcal{O}$ is replaced by the ratio of the sines of the angular radii of the shadow and of the inner edge (see Eq.~\eqref{alphash}). For $r_O\geq100\,r_{\rm ISCO}$ and $\lambda\geq5$, this ratio differs from the boundary value of $\mathcal{O}$ by less than $0.2\%$, and the ratio of the angular radii themselves by less than $0.9\%$; both differences are much smaller than the uncertainties assumed in Sec.~\ref{Sec:Forecast}. For observers much closer to the black hole, the direct image of the ISCO is formed by rays with a turning point, and Eq.~\eqref{bisco} has to be modified. For large $\lambda$, $b_{\rm ISCO}\simeq r_{\rm ISCO}$, and therefore $\mathcal{O}\propto\lambda^{-1/3}$. Table~\ref{tab:O} shows that $\mathcal{O}$ depends strongly on $\lambda$ but only weakly on $e$ (by less than $2\%$ at fixed $\lambda$). Since $\gamma$ depends only on $e$, the two observables complement each other: $\gamma$ fixes the charge, while $\mathcal{O}$ fixes the AdS radius. For an inclined observer, the image of the ISCO is no longer a circle; in this work, we restrict ourselves to the face-on case.

\begin{table*}[t]
\centering
\renewcommand{\arraystretch}{1.25}
\begin{tabular}{|c|c|c|c|c|}
\hline
$\lambda$ & $e$ & $x_{\rm ISCO}$ & $b_{\rm ISCO}$ & $\mathcal{O}$ \\
\hline
10   & 0 / 0.5 / 0.8 & 3.162 / 3.112 / 3.021 & 3.258 / 3.216 / 3.140 & 0.835 / 0.834 / 0.830 \\
\hline
30   & 0 / 0.5 / 0.8 & 4.196 / 4.140 / 4.041 & 4.372 / 4.319 / 4.225 & 0.644 / 0.641 / 0.636 \\
\hline
100  & 0 / 0.5 / 0.8 & 6.027 / 5.954 / 5.830 & 6.175 / 6.105 / 5.984 & 0.458 / 0.456 / 0.451 \\
\hline
300  & 0 / 0.5 / 0.8 & 8.555 / 8.458 / 8.294 & 8.667 / 8.571 / 8.409 & 0.326 / 0.325 / 0.321 \\
\hline
1000 & 0 / 0.5 / 0.8 & 12.68 / 12.54 / 12.31 & 12.76 / 12.62 / 12.39 & 0.222 / 0.221 / 0.218 \\
\hline
\end{tabular}
\caption{The ISCO radius $x_{\rm ISCO}=r_{\rm ISCO}/\sqrt{M}$, the radius $b_{\rm ISCO}$ of the direct image of the ISCO for a face-on observer, and the observable $\mathcal{O}=b_c/b_{\rm ISCO}$, for several values of $\lambda$ and $e$.}
\label{tab:O}
\end{table*}

\subsection{Parameter space and forecast setup}

A regular event horizon exists only for $e<e_{\rm ext}(\lambda)$, where $e_{\rm ext}$ corresponds to the extremal black hole, $f=f'=0$. Solving these two conditions gives the extremal curve in the parametric form
\begin{equation}
e_{\rm ext}^2=\frac{w(4-w)}{3}\ ,\qquad \lambda^2=\frac{3w^2}{2(1-w)}\ ,\qquad 0<w<1\ ,
\label{eext}
\end{equation}
where $w=r_{\rm ext}^2/M$. For example, $e_{\rm ext}\simeq0.99995$ for $\lambda=100$. We restrict ourselves to $\lambda\geq5$, for which all rays forming the direct image of the ISCO have $b<l$ and thus reach the asymptotic region. We adopt a uniform prior on $e\in[0,\,0.999\,e_{\rm ext}(\lambda))$ and a uniform prior on $\log_{10}\lambda\in[\log_{10}5,\,3]$.

Since $\gamma$ and $\mathcal{O}$ have not yet been measured, we perform a forecast. Future high-resolution VLBI facilities, such as the next-generation Event Horizon Telescope (ngEHT) and the proposed Black Hole Explorer (BHEX), aim to improve measurements of photon-ring structures and to provide new opportunities to probe strong-field gravity~\cite{Johnson2023ngEHT,Lupsasca2024BHEX}. Motivated by these developments, we consider illustrative fractional uncertainties of $5\%$ and $10\%$ for the two observables. These assumed uncertainties are not interpreted as forecasts for any specific observational facility.
For a chosen fiducial model $(e_\ast,\lambda_\ast)$, we take the mock data $\mathbf{D}=(\gamma_\ast,\mathcal{O}_\ast)$ with equal fractional uncertainties, $\sigma_\gamma/\gamma_\ast=\sigma_{\mathcal{O}}/\mathcal{O}_\ast=\sigma$. The likelihood is Gaussian,
\begin{equation}
\ln\mathcal{L}(e,\lambda)=-\frac{1}{2}\left[\frac{\left(\gamma(e)-\gamma_\ast\right)^2}{\sigma_\gamma^2}+\frac{\left(\mathcal{O}(e,\lambda)-\mathcal{O}_\ast\right)^2}{\sigma_{\mathcal{O}}^2}\right],
\label{likelihood}
\end{equation}
and the posterior is sampled with the affine-invariant ensemble sampler \texttt{emcee}~\cite{emcee}. As an illustrative measure of the difference from the 4D Schwarzschild benchmark, we quote the Gaussian separation $Z_\gamma=|\gamma_\ast-\pi|/\sigma_\gamma$. It is neither a Bayes factor nor a significance derived from actual observations, and it does not constitute a test against all four-dimensional black holes.

\section{Results and discussions
\label{Sec:Forecast}}

\begin{figure*}[t]
\centering
\includegraphics[width=0.95\linewidth]{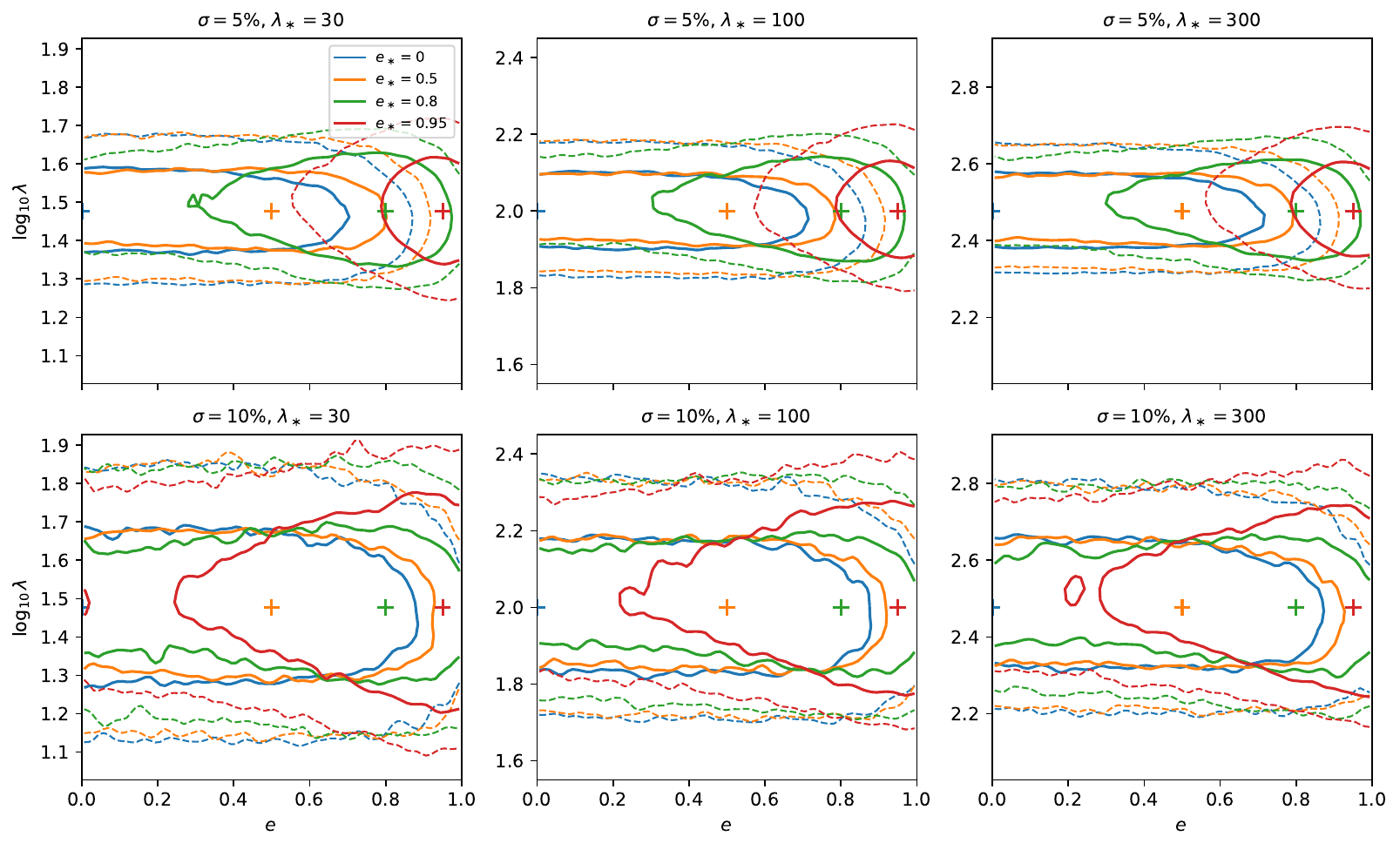}
\caption{Forecast posteriors in the $(e,\log_{10}\lambda)$ plane for $\sigma=5\%$ (top) and $\sigma=10\%$ (bottom) and for $\lambda_\ast=30$, $100$ and $300$ (left to right). Solid and dashed curves show the 68\% and 95\% credible regions, and crosses mark the fiducial models.}
\label{fig:contours}
\end{figure*}

\begin{figure*}[t]
\centering
\includegraphics[width=\linewidth]{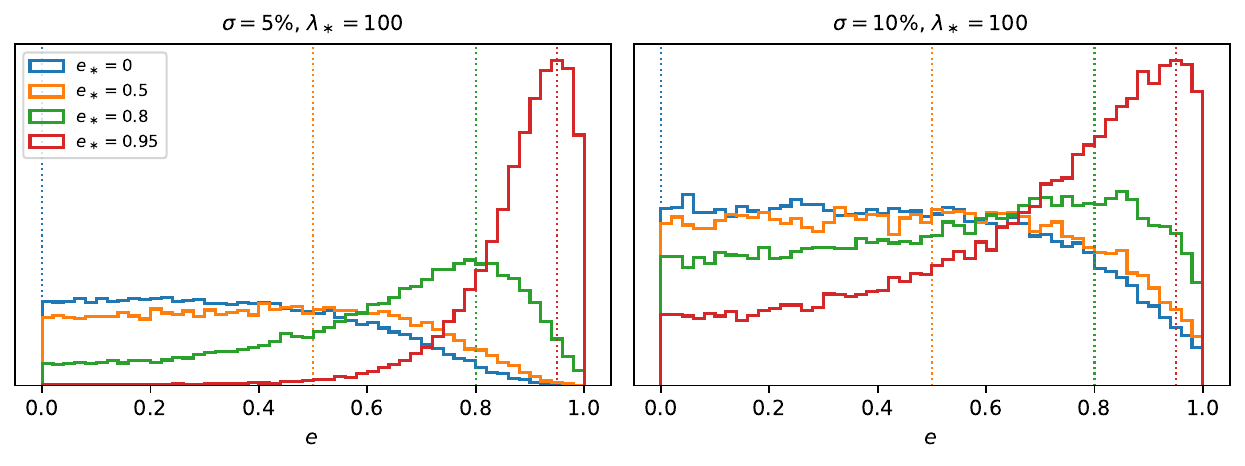}
\caption{Marginalized posteriors of the charge parameter $e$ for $\lambda_\ast=100$ and $\sigma=5\%$ (left) and $10\%$ (right). Dotted lines mark the fiducial values $e_\ast$.}
\label{fig:epost}
\end{figure*}

\begin{figure*}[t]
\centering
\begin{subfigure}{0.48\textwidth}
\centering
\includegraphics[width=\linewidth]{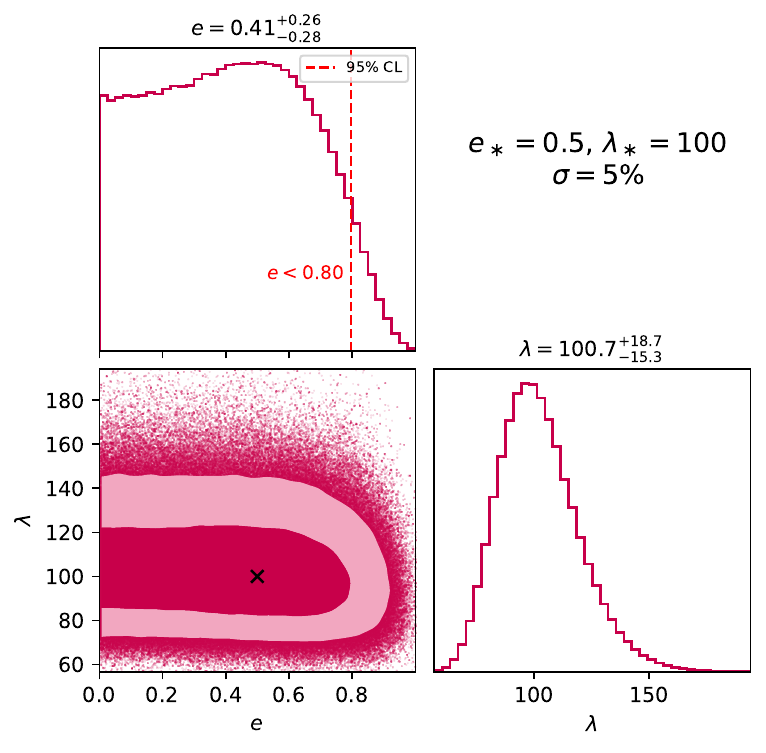}
\caption{$e_\ast=0.5$}
\end{subfigure}
\hfill
\begin{subfigure}{0.48\textwidth}
\centering
\includegraphics[width=\linewidth]{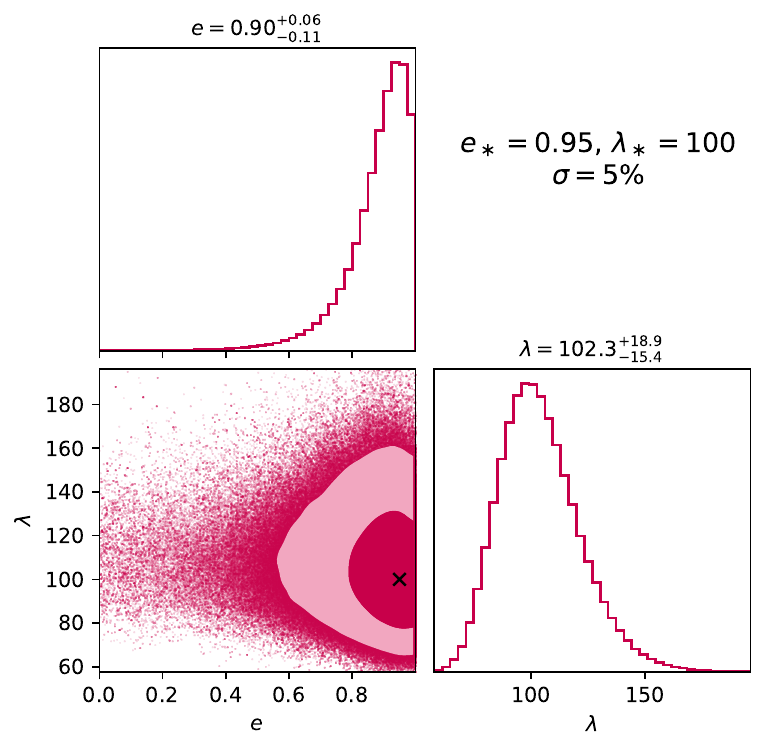}
\caption{$e_\ast=0.95$}
\end{subfigure}
\caption{Posterior distributions of $e$ and $\lambda$ for $\lambda_\ast=100$ and $\sigma=5\%$. Dark and light regions show the 68\% and 95\% credible regions, dots show the MCMC samples, and crosses mark the fiducial models. The titles give the medians and the 68\% intervals; the dashed line in panel (a) marks the 95\% upper bound on $e$.}
\label{fig:corner}
\end{figure*}

We now apply the forecast strategy described in Sec.~\ref{Sec:Obs}. We consider the fiducial models $e_\ast\in\{0,\,0.5,\,0.8,\,0.95\}$ and $\lambda_\ast\in\{30,\,100,\,300\}$, with fractional uncertainties of $5\%$ and $10\%$ on both observables. We use noiseless mock data, $\mathbf{D}=\left(\gamma(e_\ast),\,\mathcal{O}(e_\ast,\lambda_\ast)\right)$, so that the maximum of the likelihood coincides with the fiducial model and the width of the posterior reflects the assumed measurement uncertainties. For each model, we run 32 walkers for $8\times10^4$ steps and discard the first $5\times10^3$ steps as burn-in. The integrated autocorrelation time is $\tau\simeq37$--$45$ steps, i.e., every chain is longer than $1600\,\tau$, and the mean acceptance fraction lies between 0.67 and 0.70. The results for $\lambda_\ast=100$ are summarized in Table~\ref{tab:forecast}. The posterior distributions for all models are shown in Figs.~\ref{fig:contours} and~\ref{fig:epost}, and the full posteriors for two representative models are shown in Fig.~\ref{fig:corner}.

\begin{table}[t]
\centering
\renewcommand{\arraystretch}{1.35}
\begin{tabular}{|c|c|c|c|c|c|c|}
\hline
 & \multicolumn{3}{c|}{$\sigma=5\%$} & \multicolumn{3}{c|}{$\sigma=10\%$} \\
\hline
$e_\ast$ & $e$ & $\lambda$ & $Z_\gamma$ & $e$ & $\lambda$ & $Z_\gamma$ \\
\hline
0    & $<0.74$ & $99.4^{+18.6}_{-15.1}$ & $5.9$ & $<0.88$ & $100.8^{+42.4}_{-28.0}$ & $2.9$ \\
0.5  & $<0.80$ & $100.7^{+18.7}_{-15.3}$ & $5.5$ & $<0.90$ & $102.1^{+42.9}_{-28.3}$ & $2.7$ \\
0.8  & $0.68^{+0.17}_{-0.33}$ & $102.4^{+19.0}_{-15.6}$ & $4.6$ & $0.56^{+0.29}_{-0.36}$ & $105.4^{+44.2}_{-29.2}$ & $2.3$ \\
0.95 & $0.90^{+0.06}_{-0.11}$ & $102.3^{+18.9}_{-15.4}$ & $3.4$ & $0.72^{+0.20}_{-0.40}$ & $107.7^{+45.4}_{-29.8}$ & $1.7$ \\
\hline
\end{tabular}
\caption{Forecast constraints for $\lambda_\ast=100$. For $e_\ast=0$ and $0.5$, the posterior of $e$ yields only an upper bound (95\%); otherwise the median and the 68\% interval are given. The last column for each $\sigma$ is the Gaussian separation $Z_\gamma$ from the 4D Schwarzschild value $\gamma=\pi$. The results for $\lambda_\ast=30$ and $300$ are nearly identical, with $\delta\lambda/\lambda=19\%$ and $16\%$ ($\sigma=5\%$), respectively.}
\label{tab:forecast}
\end{table}

\paragraph*{AdS radius.}
The AdS radius is recovered with a relative  uncertainty of $16$--$19\%$ for $\sigma=5\%$ and $33$--$40\%$ for $\sigma=10\%$, depending only weakly on $\lambda_\ast$. This follows from the dependence of $\mathcal{O}$ on $\lambda$: for large $\lambda$, $\mathcal{O}\propto\lambda^{-1/3}$, and a fractional error $\sigma$ on $\mathcal{O}$ translates into $\delta\lambda/\lambda\simeq3\sigma$, i.e., $15\%$ and $30\%$, respectively. The correlation coefficient between $e$ and $\log_{10}\lambda$ is small, between $-0.15$ and $-0.05$, which confirms that the constraints from the two observables are nearly orthogonal in the parameter space.

\paragraph*{Charge.}
The charge parameter is much more weakly constrained, as shown in Figs.~\ref{fig:epost} and~\ref{fig:corner}. For $e_\ast=0$ and $0.5$, the posteriors are nearly identical and yield only the upper bounds $e<0.74$ and $e<0.80$ (95\%) for $\sigma=5\%$. There are two reasons for this: over the whole range $0\leq e<1$, $\gamma$ changes by only $\approx20\%$, and near $e=0$ it depends on the charge only quadratically, $\gamma\simeq\pi\sqrt{2}\left(1-3e^2/32\right)$. For $e_\ast=0.8$, the posterior peaks near the fiducial value but still extends down to $e=0$. A non-zero charge is detected only close to extremality: for $e_\ast=0.95$ and $\sigma=5\%$, we find $e=0.90^{+0.06}_{-0.11}$ and $e>0.66$ at 95\% confidence. For $\sigma=5\%$ and $e_\ast>0$, the fiducial values lie within the 68\% intervals. For $\sigma=10\%$, the charge is only weakly constrained for all fiducial models.

\paragraph*{Comparison with 4D black holes.}
Since $\gamma>3.628$ for every 5D charged AdS BH, whereas $\gamma=\pi$ for the 4D Schwarzschild BH and $\pi/\sqrt{2}\leq\gamma\leq\pi$ for the 4D Reissner--Nordstr\"om BH, the subring exponent alone can, in principle, separate the 5D model from these 4D black holes, independently of $\lambda$. For $\sigma=5\%$, the separation from the Schwarzschild value is $Z_\gamma=3.4$--$5.9$, the smallest value corresponding to the near-extremal model $e_\ast=0.95$, for which $\gamma$ is closest to $\pi$. For $\sigma=10\%$, it decreases to $Z_\gamma=1.7$--$2.9$. Since $\gamma\leq\pi$ for the 4D Reissner--Nordstr\"om BH, the separation from it is even larger. These numbers illustrate the sensitivity of $\gamma$ under the assumed Gaussian uncertainties; they do not exclude other four-dimensional geometries, which we have not considered.

\paragraph*{Complementarity of the observables.}
To illustrate the role of each observable, we repeat the analysis for $e_\ast=0.8$, $\lambda_\ast=100$ and $\sigma=5\%$ using only one of them. With $\gamma$ alone, the posterior of $\lambda$ coincides with the prior (the 68\% interval extends from $\lambda\approx12$ to $\lambda\approx430$), while the constraint on $e$ is unchanged. With $\mathcal{O}$ alone, $e$ is unconstrained, its posterior again reproducing the prior ($e=0.50\pm0.34$), while $\lambda$ is recovered with the same $17\%$ precision as in the joint analysis. Hence, $\gamma$ carries the information on the charge and controls the comparison with 4D black holes, whereas $\mathcal{O}$ constrains the AdS radius; both are needed to constrain the full parameter space.

\paragraph*{Limitations.}
This forecast is idealized. We assume that the observables $\gamma$ and $\mathcal{O}$ can be measured with the quoted fractional precision by a face-on observer in the asymptotic region, and that their uncertainties are Gaussian and uncorrelated. We neglect astrophysical systematics, such as the emission profile of the source and possible departures of its inner edge from the ISCO, as well as the effects of inclination. Moreover, the constraints refer to the dimensionless parameters $e$ and $\lambda$; converting them into a physical charge or AdS length requires relating the five-dimensional mass parameter to a dimensional scale. The results should therefore be regarded as an estimate of the theoretical sensitivity of these geometric observables to the parameters of the 5D charged AdS BH.

\section{Conclusion} \label{Sec:Conc}

In this work, we investigated the photon dynamics, the shadow, and the ray-tracing signatures of a five-dimensional charged AdS black hole. We also introduced two mass-independent geometric observables and performed an MCMC forecast to assess how well they can constrain the black hole charge and the AdS radius.

More specifically, first, using the Euler–Lagrange and null geodesic equations, photon orbits have been investigated around a 5D charged AdS BH. In addition, key quantities of this BH (shadow, photon sphere, and event horizon radii) have been calculated. As the charge parameter $Q$ increases, all of these radii decrease. However, as the AdS radius $l$ increases, all of them except the photon sphere radius increase, because the photon sphere radius does not depend on the AdS radius (see Fig.~\ref{fig:function_r}). Also, we have plotted ray-tracing images (Figs.~\ref{b_func} and \ref{ray1}).
The ray-tracing results show that increasing the charge parameter reduces the critical impact parameter and changes photon trajectories near the black hole. 
For example, as shown in Table~\ref{table1}  and Fig.~\ref{F_b1}, when the charge parameter changes from 0.2 to 0.8, the critical impact parameter $b_c$ decreases from 2.7153 to 2.6046, and the photon ring interval shifts from $(2.715< b < 2.717)$ to $(2.604 < b < 2.607)$.

Finally, we have introduced two dimensionless observables to characterize the 5D charged AdS BH without relying on an absolute mass scale. This approach avoids the ambiguity associated with relating the five-dimensional mass parameter, which has dimensions of length squared, to black hole masses inferred from four-dimensional observations. Moreover, it allows us to investigate the spacetime geometry without relying on the absolute apparent size of the shadow, which depends on the observer's position in AdS spacetime.

The first observable is the subring exponent $\gamma$, which characterizes how rapidly successive photon subrings approach the critical curve. We have derived a closed-form expression for this quantity, Eq.~\eqref{gammaclosed}, and shown that it is independent of both the AdS radius and the observer's position. As the charge increases, $\gamma$ decreases from $\pi\sqrt{2}\simeq4.44$, approaching $2\pi/\sqrt{3}\simeq3.63$ in the asymptotically flat extremal limit. It remains above the values $\gamma\leq\pi$ corresponding to the 4D Schwarzschild and Reissner--Nordstr\"om black holes (see Table~\ref{tab:gamma}).
The second observable is the ratio $\mathcal{O}$ of the critical impact parameter to the impact parameter corresponding to the direct image of the ISCO, Eq.~\eqref{Oratio}. We have shown that stable circular orbits, and hence the ISCO, exist in this spacetime due to the AdS term, with $r_{\rm ISCO}\propto l^{1/3}$ for large $l$. Consequently, $\mathcal{O}$ is primarily sensitive to the AdS radius and only weakly dependent on the charge (see Table~\ref{tab:O}).

Using these observables, we have performed an MCMC forecast based on noiseless mock data (Sec.~\ref{Sec:Forecast}). For assumed fractional uncertainties of $5\%$, the AdS radius is constrained with a relative uncertainty of $16$--$19\%$, which increases to $33$--$40\%$ for uncertainties of $10\%$ (see Table~\ref{tab:forecast} and Fig.~\ref{fig:contours}). The charge is constrained much more weakly. For small and moderate charges, only upper bounds are obtained, e.g., $e<0.74$ and $e<0.80$ at the 95\% credible level for $e_\ast=0$ and $0.5$. The charge can be distinguished from zero only for the near-extremal fiducial model $e_\ast=0.95$ with $\sigma=5\%$, for which we find $e=0.90^{+0.06}_{-0.11}$ (68\% credible interval; see Figs.~\ref{fig:epost} and~\ref{fig:corner}). This weak sensitivity originates from the small variation of $\gamma$ with the charge, especially at small $e$, where its dependence is quadratic. At the same time, since the range of $\gamma$ for the 5D black hole does not overlap with those for the 4D Schwarzschild and Reissner--Nordstr\"om black holes, the subring exponent can, in principle, distinguish the considered 5D geometry from these 4D black holes: for $5\%$ uncertainties, the Gaussian separation from the Schwarzschild value is $Z_\gamma=3.4$--$5.9$, while for $10\%$ it reduces to $1.7$--$2.9$. We have also found that the two observables play complementary roles: $\gamma$ carries the information on the charge, whereas $\mathcal{O}$ constrains the AdS radius, and only their combination constrains both parameters.

These results rely on an idealized setup, which includes a face-on observer, Gaussian and uncorrelated uncertainties, and no astrophysical systematics, and they refer to the dimensionless parameters $e$ and $\lambda$. They should therefore be regarded as an estimate of the sensitivity of mass-independent geometric observables to the parameters of the 5D charged AdS BH. Extending the analysis to inclined observers and to more realistic emission models is an interesting direction for future work.

\bibliographystyle{apsrev4-1}
\bibliography{ref1,ref2,refnew,ref1new,ref2new}

\end{document}